\documentclass{moriond}

\def\Journal#1#2#3#4{{#1} {\bf #2}, #3 (#4)}

\def\PLB{{\em Phys. Lett.}  B}

\def\ZPC{{\em Z. Phys.} C}
\def\JINST{{\em JINST}}
\def\JHEP{{\em JHEP}}

\def\be{\begin{equation}}
\def\ee{\end{equation}}
\def\bea{\begin{eqnarray}}
\def\eea{\end{eqnarray}}

\usepackage{xspace}
\usepackage{cite}

\usepackage{amsmath}
\newcommand{\N}{\ensuremath{N}\xspace}
\newcommand{\VeN}{\ensuremath{V^{}_{\ele \N}}\xspace}
\newcommand{\VmN}{\ensuremath{V^{}_{\muon \N}}\xspace}

\newcommand{\VlN}{\ensuremath{V^{}_{\ell \N}}\xspace}
\newcommand{\mN}{\ensuremath{m_{\N}}\xspace}
\newcommand{\mW}{\ensuremath{m_{\W}}\xspace}
\newcommand{\mZ}{\ensuremath{m_{\Z}}\xspace}
\newcommand{\MT}{\ensuremath{M_\text{T}}\xspace}

\newcommand{\Mtril}{\ensuremath{M_{3\ell}}\xspace}
\newcommand{\minMOS}{\ensuremath{M_{2\ell\text{OS}}^{\min}}\xspace}
\newcommand{\W}{\ensuremath{W}\xspace}
\newcommand{\Z}{\ensuremath{Z}\xspace}
\newcommand{\WZ}{\ensuremath{WZ}\xspace}
\newcommand{\ZZ}{\ensuremath{ZZ}\xspace}
\newcommand{\ttbar}{\ensuremath{\text{t}\overline{\text{t}}}\xspace}
\newcommand{\lep}{\ensuremath{\ell}\xspace}
\newcommand{\ele}{\ensuremath{e}\xspace}
\newcommand{\muon}{\ensuremath{\mu}\xspace}
\newcommand{\GeV}{\ensuremath{\text{GeV}}\xspace}
\newcommand{\pt}{\ensuremath{p_{\text{T}}}\xspace}
\newcommand{\met}{\ensuremath{E_{\text{T}}^{\text{miss}}}\xspace}

\newcommand{\Photo}{
}

\begin{document}
\vspace*{4cm}
\title{Search for heavy neutral leptons with the CMS detector}

\author{Willem Verbeke}

\address{On behalf of the CMS Collaboration\\
Department of Physics and Astronomy,\\
Universiteit Gent, Gent, Belgium}

\maketitle\abstracts{
The smallness of neutrino masses provides a tantalizing allusion to physics beyond the standard model. Heavy neutral leptons (\N), such as hypothetical sterile neutrinos, accommodate a way to explain this observation, through the see-saw mechanism. If they exist, \N could also provide answers about the nature of dark matter, and the baryon asymmetry of the universe. A search for the production of \N at the LHC, originating from leptonic W boson decays through the mixing of \N with SM neutrinos, is presented. The search focuses on signatures with three leptons, providing a clean signal for probing the production of \N in a wide mass range never explored before at the LHC: down to 1 GeV, and up to 1.2 TeV. The sample of proton-proton collisions collected by the CMS detector throughout 2016 is used, amounting to a volume of 35.9 $\text{fb}^{-1}$. The results are presented in the plane of the mixing parameter of \N to their SM counterparts, versus their mass, and are the first such result at a hadron collider for masses below 40 GeV and the first direct result for masses above 500 GeV, more than doubling the probed mass range.}

\section{Heavy neutral leptons}
Through the observation of neutrino oscillations, it has been established that neutrinos have non-zero, albeit very small, masses. This finding, and other unexplained phenomena, such as dark matter and the baryon asymmetry of the universe, strongly hint at the existence of physics beyond the standard model (SM). The introduction of heavy neutral leptons (\N), otherwise known as sterile neutrinos, could solve several of these outstanding problems. Through the see-saw mechanism their presence could explain the smallness of neutrino masses, while a lighter \N might be a dark matter candidate, and heavier \N could simultaneously provide enough CP violation to explain the observed matter-antimatter asymmetry~\cite{Asaka:2005an,Asaka:2005pn}.

A search for \N production is performed using the CMS detector at the LHC~\cite{CMSDetector}. The data were collected in proton-proton collisions at a center of mass energy of 13 TeV, and correspond to an integrated luminosity of 35.9 $\text{fb}^{-1}$~\cite{trilepton}.

\section{Search in the trilepton final state}
\begin{figure}
\centering
\includegraphics[width=.3\textwidth]{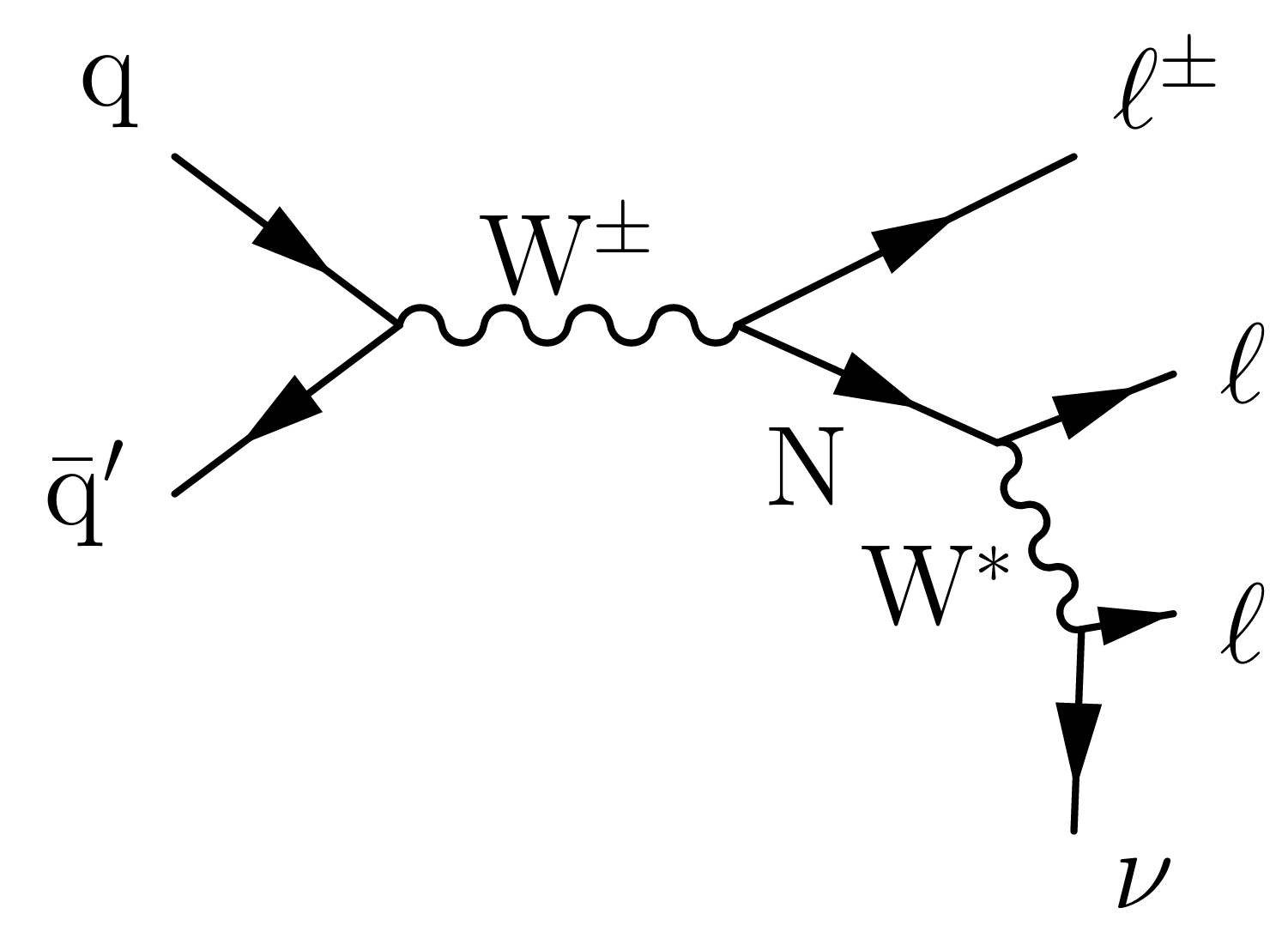}
\caption{Feynman diagram of \N production in a leptonic \W decay, leading to a final state with three charged leptons and a neutrino.}
\label{fig:diagram}
\end{figure}

This search focuses on the production of \N through the leptonic decay of W bosons,  $\W^{(*)} \to \N \lep$ ($\lep = \ele, \muon$), where \N promptly decays to $\W^{(*)}\lep$. With a subsequent decay of the last vector boson to $\lep\nu$, this leads to a final state with three leptons and a neutrino, as depicted in Fig.~\ref{fig:diagram}. The presence of the neutrino makes it impossible to reconstruct the full \N mass, and thus to hunt for a mass peak. While reconstructing the \N mass is possible in final states where the last \W decays hadronically, the trilepton final state has the distinct advantage that it facilitates the probing of very low \N masses, through the possibility to reconstruct soft leptons, down to 5 (10) \GeV in \pt for muons (electrons). 

When \VlN and \mN become very low, the \N will start getting a significant lifetime, leading to the two leptons from the \N decay being displaced w.r.t. the primary interaction vertex. In this analysis the focus lies on prompt \N decays, requiring the presence of leptons compatible with the primary interaction vertex. 

\section{Analysis strategy}
Mirroring the signal topology, events containing exactly 3 well identified electrons or muons are  retained for analysis. The SM backgrounds in this final state consists of events in which there are three genuine (prompt) leptons, and events in which one or more leptons come from jet-fragmentation or misidentified particles in the detector (nonprompt). The search looks at events without b-tagged jets, so the former category consists mainly of the \WZ, $\W\gamma^{(*)}$, \ZZ, and $\Z\gamma^{(*)}$ processes. The latter category is dominated by \ttbar and Drell-Yan events with an additional nonprompt lepton. 

Most of the aforementioned backgrounds have the trait that they tend to contain a pair of leptons with opposite sign and same flavor (OSSF), with background yields being more than two orders of magnitude larger for events containing an OSSF pair than for events without such a pair. Because of the Majorana nature of \N, its decay can violate lepton number, enabling the possibility of events without an OSSF pair. To exploit this peculiarity of the signal, events with- and without an OSSF pair are analyzed separately. 

To be sensitive to \mN both above- and below \mW the analysis is divided in two orthogonal categories, dubbed the low- and high-mass regions.
 
The low-mass region is defined by values of the trilepton invariant mass (\Mtril) below \mW, and relatively low values of the leading lepton \pt and missing transverse energy (\met). In this region leptons with \pt as low as allowed by the online triggers used to record the data are reconstructed. Because of the prodigious $\W\gamma^{(*)}$ background in this kinematic region, only events without an OSSF pair are retained here.

The high-mass region is characterized by high values of the leading lepton \pt, and higher \pt thresholds on all leptons. At higher \mN, the kinematic shapes of \N decays become easier to distinguish from SM processes, so significant sensitivity can be won by retaining, and separately analyzing, events with an OSSF pair. In order to suppress the WZ, ZZ and $\Z\gamma^{(*)}$ backgrounds in these events, those in which \Mtril, or the mass of any OSSF pair, is compatible with \mZ, are vetoed.

The minimum mass out of all lepton pairs of opposite charge (\minMOS), is strongly correlated with \mN in signal events, and on that account, both low- and high-mass regions are divided into multiple \minMOS bins, sensitive to different \mN ranges. At high \mN, the transverse mass (\MT) of the lepton not present in the pair forming \minMOS, and the \met, tends to be much larger in signal events than in SM processes. Therefore each of the \minMOS bins in the high-mass region is further subdivided into bins of \MT. To achieve enhanced distinction between signal and background, events having $\Mtril$ smaller than 100 GeV are evaluated independently in the high-mass region. 

\section{Results}
\begin{figure}
\centering
\includegraphics[width=.8\textwidth]{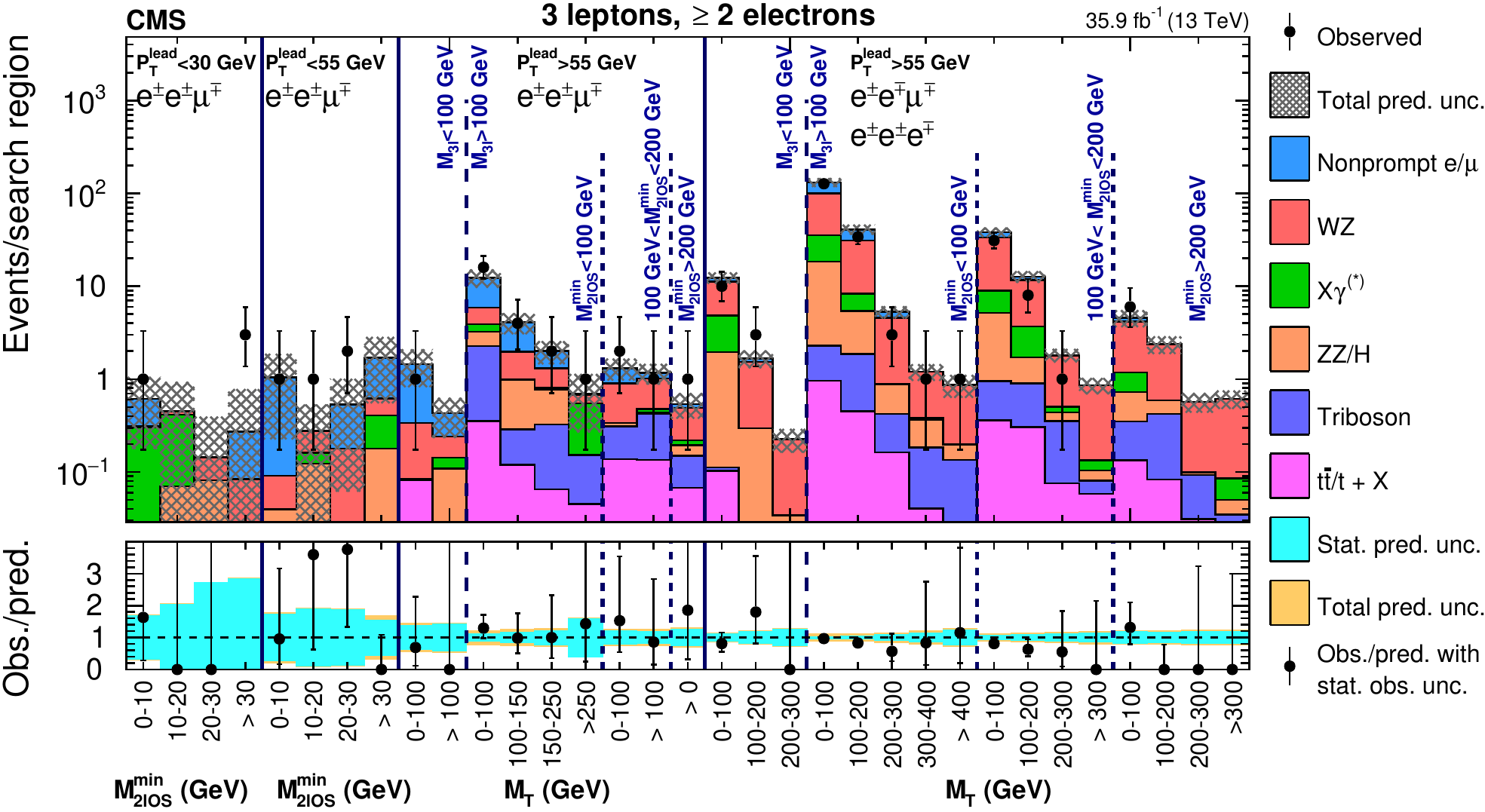}
\includegraphics[width=.8\textwidth]{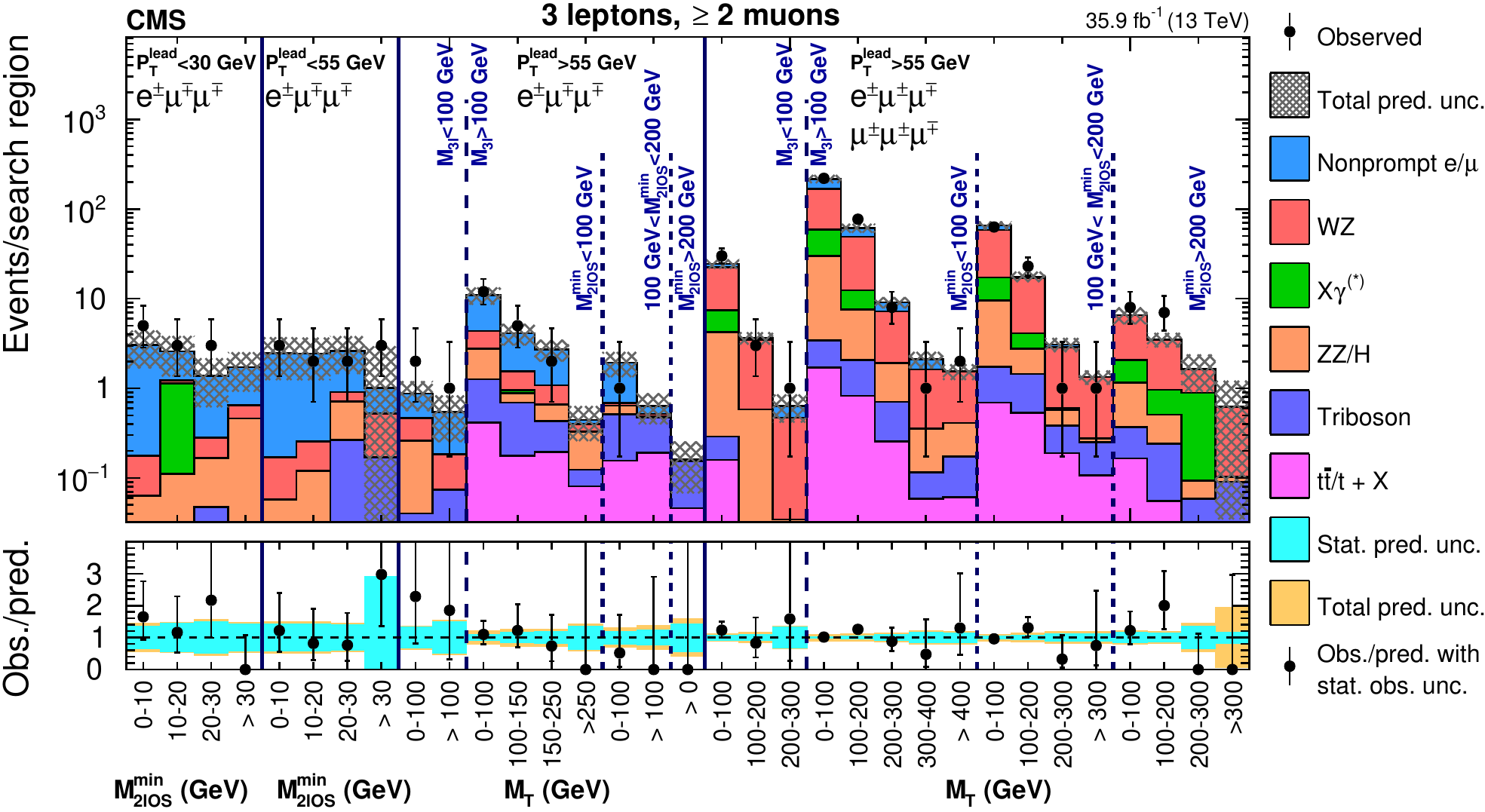}
\caption{Observed and expected yields in the search regions, for events with at least 2 electrons (top), and events with at least 2 muons (bottom). The first 8 bins of each plot correspond to the low-mass region, while the rest depict the high-mass region~\cite{trilepton}.}
\label{fig:SRplots}
\end{figure}

\begin{figure}
\includegraphics[width=.5\textwidth]{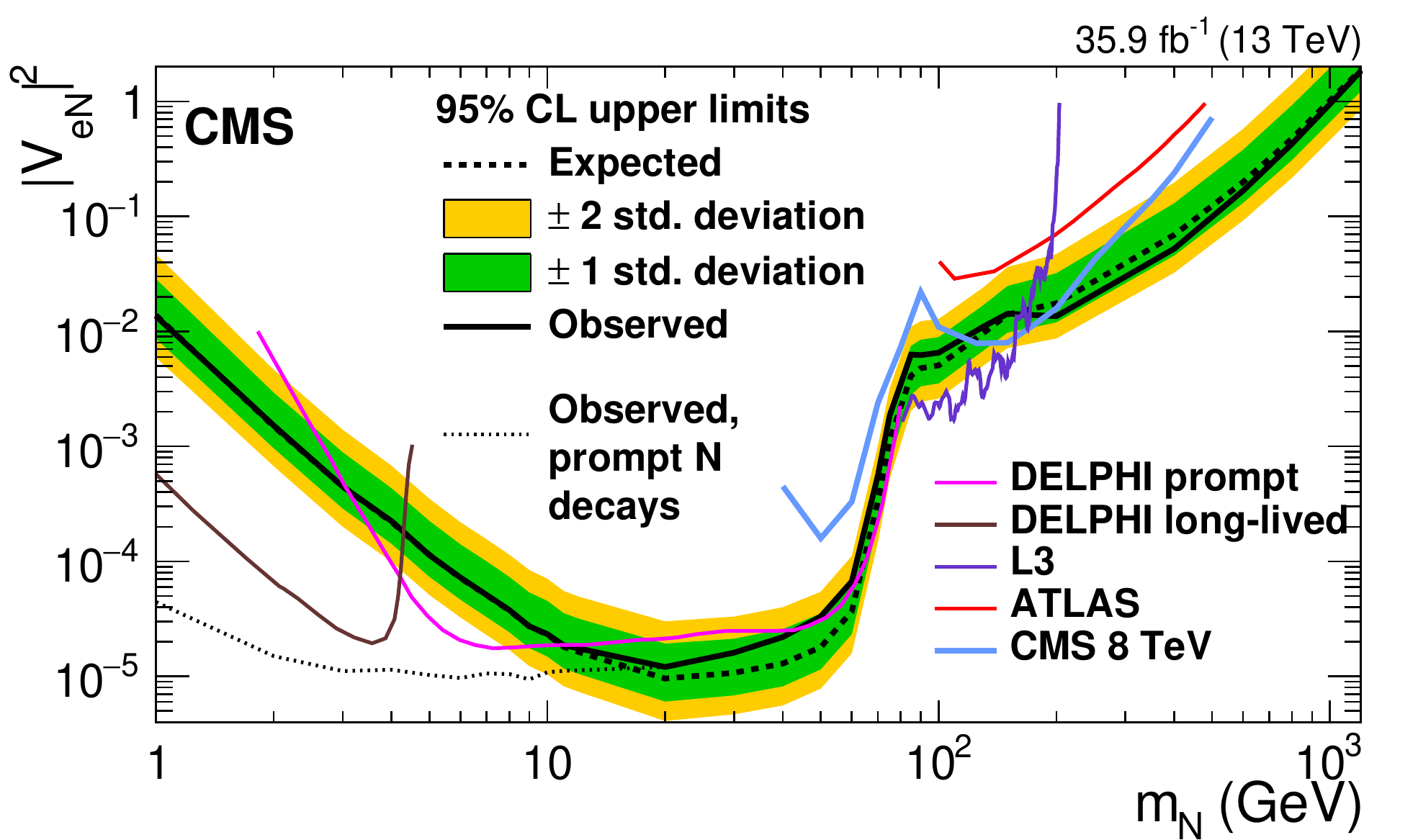}
\includegraphics[width=.5\textwidth]{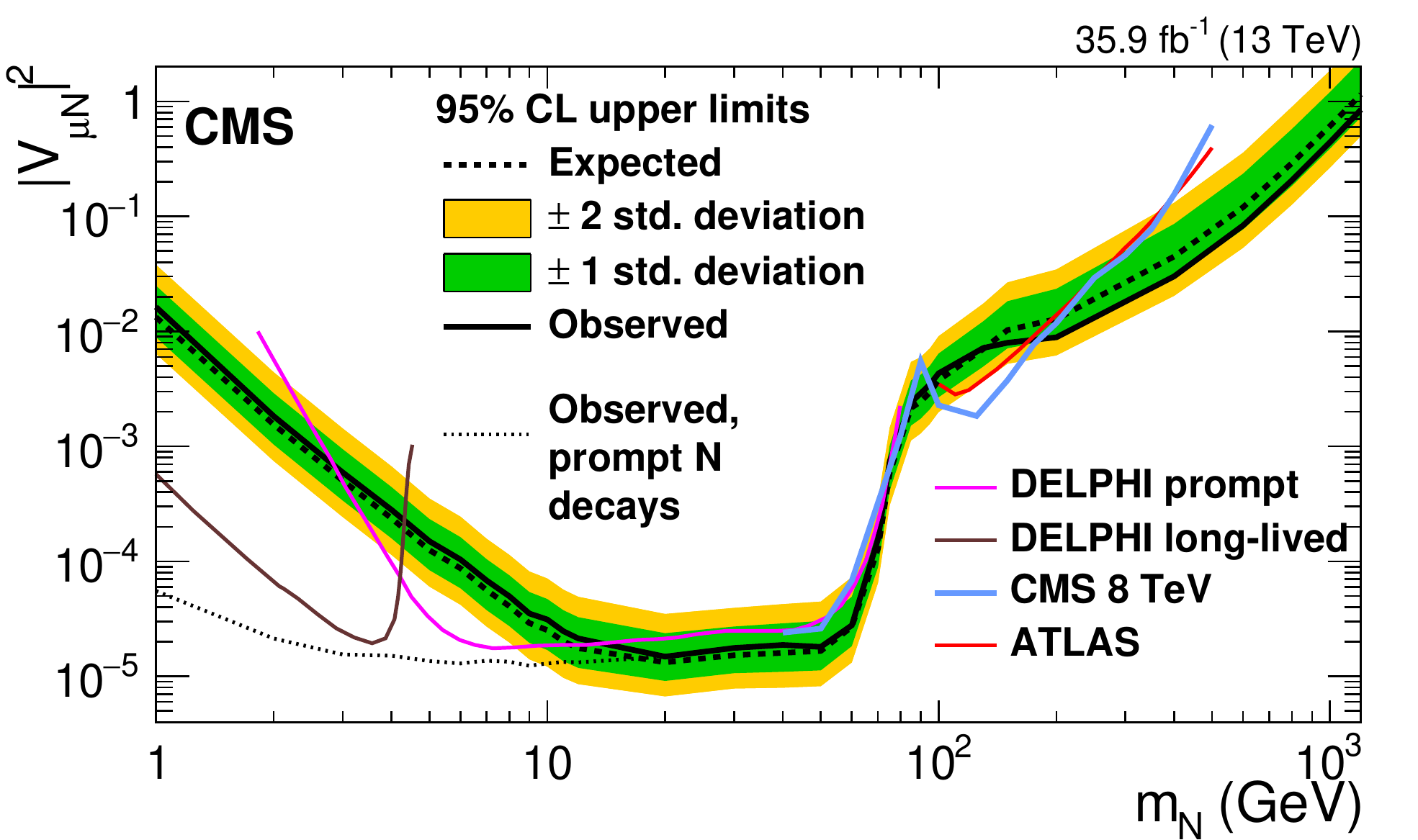}
\caption{95\% confidence upper limits on \VeN and \VmN as a function of \mN~\cite{trilepton}.}
\label{fig:limits}
\end{figure}

In order to probe \VeN and \VmN separately, events with 2 or more electrons, and 2 or more muons are respectively considered. The observed and expected yields in these categories, together with the definition of each of the search regions can be found in Fig.~\ref{fig:SRplots}. No statistically significant excess above the SM expectation is observed, and upper limits at the 95\% confidence level on \VeN and \VmN, displayed in Fig.~\ref{fig:limits}, are set as a function of \mN. 

The limits attained with this search can be seen to surpass those from DELPHI~\cite{DELPHI} for masses above about 15 \GeV, the first improvement since the LEP era, and the first time this range is probed at a hadron collider. At high masses, the results improve upon the 8 TeV results from CMS~\cite{CMS8TeVmumu,CMS8TeVee} and ATLAS~\cite{ATLAS8TeV}, and limits are set up to $\mN = 1.2$ $\text{TeV}$, more than doubling the mass range that directly probed before this result. The limits at very low \mN deteriorate as \mN decreases, because the lifetime of \N becomes significant in this region, and the analysis starts losing efficiency in the reconstruction of signal leptons.

\section{Outlook}
An analysis probing the production of \N in the trilepton final state was carried out, using the proton-proton collision data recorded by CMS in 2016, corresponding to an integrated luminosity of 35.9 $\text{fb}^{-1}$~\cite{trilepton}. The analysis improves upon the state of the art upper limits in a large \mN range, and explores extensive previously uncharted phase space. As of now, the statistical uncertainty is the dominant source of uncertainty across the entire search. In 2017 CMS collected a data-volume larger than what was accumulated in 2016, and in 2018 a similar volume of data is expected to be recorded. In addition searches for \N can be performed in several other final states, each optimal in a different region of phase space. The prompt same-sign dilepton + jets final state, which due to  its large branching fraction is very sensitive at high masses, has recently been analyzed~\cite{koreans}, while the final states with displaced leptons and jets, targeting low \mN and \VlN, remain open for inspection. With this wealth of both new data and final states, ready to be exploited, the sensitivity of searches for \N at the LHC is expected to considerably increase over the next few years.

\section*{References}

\bibliography{auto_generated}

\end{document}